\documentstyle[twocolumn]{jpsj}

\def\beq{\begin{equation}}
\def\eeq{\end{equation}}
\def\beqa{\begin{eqnarray}}
\def\eeqa{\end{eqnarray}}
\def\beqan{\begin{eqnarray*}}
\def\eeqan{\end{eqnarray*}}

\def\runtitle{Temperature Dependence of Spin and Bond Ordering in a Spin-Peierls System}
\def\runauthor{Hiroaki {\sc Onishi} and Seiji {\sc Miyashita}}

\title{\bf
Temperature Dependence of Spin and Bond Ordering \\
in a Spin-Peierls System
}

\author{
Hiroaki {\sc Onishi} and Seiji {\sc Miyashita}$^1$
}

\inst{
Department of Earth and Space Science, Graduate School of Science, Osaka University,\\
Machikaneyamacho 1-1, Toyonaka 560-0043\\
$^1$Department of Applied Physics, University of Tokyo,
    Hongo 7-3-1, Bunkyoku, Tokyo 113-0033}

\recdate{September 6, 2000}
\abst{
We investigate thermodynamic properties of
a one-dimensional $S=1/2$ antiferromagnetic Heisenberg model coupled to a lattice distortion
by a quantum Monte Carlo method.
In particular we study how spin and lattice dimerize
as a function of the temperature,
which gives a fundamental process of the spin-Peierls transition in higher dimensions.
The degree of freedom of the lattice is taken into account adiabatically
and the thermal distribution of the lattice distortion is obtained by the thermal bath algorithm.
We find that the dimerization develops as the temperature decreases
and it converges to the value of the dimerization of the ground state at $T=0$.
Furthermore we find that
the coupling constants of spins fluctuate quite largely  at high temperature
and there thermodynamic properties deviate from those of the uniform chain.
Doping of non-magnetic impurities causes
cut of the chain into short chains with open boundary.
We investigate thermodynamic properties of open chains
taking relaxation of the lattice into consideration.
We find that
strong bonds locate at the edges
and a defect of the bond alternation appears in the chain
with an odd number of sites,
which causes enhancement of a staggered magnetic order.
We find a spreaded staggered structure
which indicates that the defect moves diffusively in the chain
even at very low temperature.
}

\kword{
spin-Peierls system,
lattice fluctuation,
quantum Monte Carlo method,
bond alternation,
magnetic susceptibility,
impurity
}

\begin{document}
\sloppy
\maketitle

\section{Introduction}

The one-dimensional $S=1/2$ antiferromagnetic Heisenberg chain
is unstable to the static lattice dimerization
which is well known as the spin-Peierls instability.
The nature of the spin-Peierls transition has been investigated extensively.
Experimentally several compounds
such as TTF-CuBDT~\cite{exp1} and CuGeO$_3$~\cite{exp2}
have been found out to show the spin-Peierls transition.
In particular since the discovery of the inorganic spin-Peierls compound CuGeO$_3$,
extensive experiments have been performed
such as the neutron scattering measurement~\cite{exp-ns}
and the investigation of the effect of substitution of impurities,~\cite{exp-imp1}
which is difficult in the organic compounds.
It has been found that
spin-Peierls order is suppressed
and antiferromagnetic order is favored
with doping impurities.~\cite{exp-imp2}
Several phases appear according to the concentration of the impurity
and coexistence of spin-Peierls and antiferromagnetic states
has been observed.~\cite{exp-imp3,exp-imp4}
On the other hand, theoretically
the adiabatic lattice distortion has been considered
and the property of the ground state has been investigated
by a bosonization theory.~\cite{Cross-Fisher,Nakano-Fukuyama1,Nakano-Fukuyama2}
It is known that
an arbitrarily small spin-phonon coupling causes spontaneous lattice dimerization
in the ground state
because
the magnetic energy gain of the spin interaction in the dimerized state,
which is proportional to $u^{4/3}$, is larger than the restoring elastic energy $\propto u^2$,
where $u$ is an amount of the dimerization.
However, it is difficult to estimate thermodynamic properties exactly at finite temperature.
The thermal fluctuation of the lattice distortion has been treated self-consistently
by a mean field theory~\cite{Pytte},
a Monte Carlo method~\cite{MC-self} and a DMRG method.~\cite{DMRG-self}
In a mean field picture,
the lattice has the uniform configuration above the transition temperature
and the system has the same property as that of the uniform Heisenberg system.
However, the effect of the thermal fluctuation of the lattice distortion is not clear.
The quantum lattice fluctuation is considered to cause different effect
from that of the adiabatic lattice distortion.
It has been predicted that
in case of a finite frequency
a critical spin-phonon coupling exists
for the spin-Peierls transition.~\cite{ph-c1,ph-c2,ph-c3,ph-c4}
Thermodynamic properties have been investigated
by a quantum Monte Carlo method
taking the thermal fluctuation of the quantum phonon into consideration.~\cite{ph-1,ph-2}
It has been pointed out that
the thermal fluctuation of the quantum phonon
causes different thermodynamic properties in the spin-Peierls system
from those in the static uniform chain
even at high temperature.

The effect of the inhomogeneity of the lattice has been studied extensively
in quantum spin systems.
It has been found that peculiar magnetic structures appear
in the open chain due to the edge effect.~\cite{imp-EA,imp-L}
For example in the $S=1/2$ uniform Heisenberg antiferromagnet with open boundary,
a spreaded staggered magnetic structure appears in the odd chain,
while a one-node structure appears in the even chain.~\cite{imp-nishino1}
In the bond-alternating lattice,
the defect of the alternation causes a localized magnetic structure
around the defect.~\cite{imp-nishino2}
In the spin-Peierls system
the change from the uniform chain to the bond-alternating chain
occurs spontaneously.
Thus it is an interesting problem to study the magnetic structure
in the open chain of the spin-Peierls system with the lattice relaxation.
The effect of the lattice relaxation in the open chain has been studied
in a self-consistent way.~\cite{imp-sim1,imp-sim2,imp-sim3}
There strong bonds locate at the edges.
Furthermore, a localized structure of a defect of the bond alternation appears
in the chain with an odd number of sites
and a local staggered magnetization appears around it.
Impurity problem has also been studied by the bosonization theory.~\cite{imp-boson}

In this paper
we study the effect of the thermal fluctuation of the adiabatic lattice distortion.
In order to investigate thermodynamic properties,
we use a quantum Monte Carlo loop algorithm
with continuous imaginary time (LCQMC).~\cite{loop-1,loop-2,loop-c}
This algorithm is useful to explore the property at low temperature.
Besides, there is no need to extrapolate with respect to the Trotter number.
We extend the standard LCQMC to the system with the lattice distortion.
The thermal distribution of the lattice distortion is obtained by the thermal bath algorithm.
In \S2
model and method are explained.
As a reference to the study of the thermal fluctuation of the lattice distortion,
we review the property of the dimerized ground state
investigated by an exact diagonalization method (ED).
In \S3
we investigate thermodynamic properties of the system with the periodic boundary.
We show how the dimerization develops
as a function of the temperature.
In \S4
we investigate properties of open chains,
which is relevant to the effect of doping of non-magnetic impurities
because it causes cut of the chain into short chains with open boundary.
In \S5
summary and discussion are given.

\section{Model and Method}

The one-dimensional $S=1/2$ antiferromagnetic Heisenberg model
coupled to the adiabatic lattice distortion is given by
\beqa
H &=& \sum_{i=1}^N J_i(\tilde{\Delta}_i)\mbox{\boldmath $S$}_i\cdot\mbox{\boldmath $S$}_{i+1}
  + {\tilde{k}\over2}\sum_{i=1}^N \tilde{\Delta}_i^2, \\
\tilde{\Delta}_i &=& \tilde{u}_i-\tilde{u}_{i+1},
\eeqa
where $\tilde{u}_i$ is the lattice displacement of the $i$-th site from its equilibrium position
and $\tilde{\Delta}_i$ is the lattice distortion owing to the lattice displacement.
The distortion of the exchange coupling is assumed to be proportional to the lattice distortion
\beq
J_i(\tilde{\Delta}_i)=J(1+\alpha \tilde{\Delta}_i).
\eeq
Hereafter, for the simplicity of notation,
we use the distortion of the interaction $\alpha \tilde{\Delta}_i$ as a variable,
namely, $\Delta_i=\alpha \tilde{\Delta}_i$ and also scaled constant $k=\tilde{k}/\alpha^2$,
which leads
\beq
\label{eq:hamsp}
H = J\sum_{i=1}^N (1+\Delta_i)\mbox{\boldmath $S$}_i\cdot\mbox{\boldmath $S$}_{i+1}
  + {k\over2}\sum_{i=1}^N \Delta_i^2.
\eeq
When we consider the pure system,
the periodic boundary condition is adopted in \S3.
In order to investigate the impurity effect,
we consider open chains in \S4.
We set the uniform exchange coupling $J=1$
and take it as the unit of the energy.

We investigate thermodynamic properties of this model
by a Monte Carlo method.
The degree of freedom of the lattice is taken into account
and the thermal distribution of the lattice distortion is obtained.
The procedure to update the system is as follows.
We start with an arbitrary spin and bond configuration.
We update the spin configuration with the fixed bond configuration by the standard LCQMC.
Then we update the bond configuration with the fixed spin configuration
by the thermal bath algorithm.
For convenience of the simulation,
change of the bond distortion $\Delta_i$ is discretized by a small unit $\Delta_{\rm unit}$,
which is an approximation of continuous change.
At each update of $\Delta_i$,
three states are examined as the next value
as depicted in Fig.~\ref{fig:site}.
We consider a motion of the $i$-th site $u_i$.
The site moves right or left with the unit or stays the present position,
that is, $\delta u_i=\pm\Delta_{\rm unit}$ or $0$.
According to this site movement,
the bond distortions on both sides of the site are changed as
\beqa
\left\{
\begin{array}{lllll}
\Delta_{i-1} &\!\!\!\!\! \rightarrow 
  &\!\!\!\!\! \Delta_{i-1} &\!\!\!\!\! - &\!\!\!\!\! \delta u_i \\
\Delta_{i}   &\!\!\!\!\!\! \rightarrow
  &\!\!\!\!\! \Delta_{i}   &\!\!\!\!\! + &\!\!\!\!\! \delta u_i.
\end{array}
\right.
\eeqa
In this case the total length of the lattice does not change: \,
\beq
\label{eq:length}
\sum_{i=1}^N \Delta_i = 0.
\eeq
In realistic materials, the total length of the lattice changes
with the temperature and also with the magnetic field.
These effects are studied as the magneto-elastic property or the Invar problem,
which are interesting topics in magnetism.
However, in this paper we concentrate ourselves on the effect of
dimering due to the alternate deformation of the lattice
as an essential mechanism of the spin-Peierls phenomena.
If we allow the total length change,
the system shows the dimerization
but we found that
the total length decreases in order to gain the magnetic energy,
i.e., to increase the amplitude of the interaction $J$.
Thus we introduce the restriction eq.~(\ref{eq:length}) here.
The detail study of the effect of this restriction will be reported elsewhere.
In case of the open chain we fix the edge sites
in order to keep the total length of the lattice.
These two updates, namely,
LCQMC for the spin configuration and the bond-update,
are applied alternately.

In the simulation we set the unit of the bond distortion $\Delta_{\rm unit}=0.02$
and provide the cut off $\Delta_{i}/\Delta_{\rm unit}=-30 \sim 30$
to restrict the exchange coupling to be antiferromagnetic ($J_{i}=0.4 \sim 1.6$).
Starting from spin all up configuration and uniform bond configuration,
initial $10^5$ Monte Carlo steps (MCS) are discarded to obtain thermal equilibrium state
and then in most cases
$10^6$ MCS are performed to sample the data of physical properties.
In case of the open chain
$10^7$ MCS are performed to obtain good convergence of the data.
The simulation through the MCS is divided into 10 bins
and the errorbar is estimated from the standard deviation of the data of the bins.

\subsection{Dimerized ground state}

Before considering the thermal fluctuation of the lattice distortion,
it would be useful to discuss ground state properties of the dimerized chain with
\beq
\Delta_i=(-1)^i\delta,
\eeq
in order to investigate ground state properties of the spin-Peierls system. 
Let us check the well-known relation $\delta_0\propto k^{-3/2}$.
The dimerization amplitude in the ground state $\delta_0$
is determined by ED for values of $\delta$.
In Fig.~\ref{fig:dele}
we show the $\delta$ dependence of the change of the ground state energy,
$\Delta E_{\rm g}(\delta)$, for $N=20$.
There is a value of $\delta$ which gives a minimum value of $\Delta E_{\rm g}$,
and the dimerized ground state is realized.
In Fig.~\ref{fig:del0}
we show the $k$ dependence of $\delta_0$ for $N=20$.
There we find the relation $\delta_0\propto k^{-3/2}$ clearly.
It is a noteworthy fact that
$\delta_0$ is affected by the finite size effect for large $k$.
Actually,
$\delta_0$ is found to be zero for large $k$ in case of a small system size,
e.g., $\delta_0$ is zero for about $k>3.5$ in the case of $N=20$.
We have checked that
the values of $\delta_0$ for $k=1.0$ and $2.0$,
which will be used in the next section,
do not change among the cases of $N=16,$ $20$ and $24$.
Thus we confirmed the previously known properties of the distortion in the ground state.

As shown in Fig.~\ref{fig:dele},
it should be noted that
the amount of the energy gain due to the dimerization $\Delta E(k)$ is rather small
and also that
the dependence of $\Delta E_{\rm g}$ on $\delta$, namely,
the effective potential for the distortion,
is very weak in the region where $\Delta E_{\rm g}<0$.
Thus even at low temperature, e.g., at $T=0.1$ for $k=1.0$,
large fluctuation of the lattice is expected,
which will be seen in the next section.

\section{Thermodynamic Properties}

\subsection{Bond fluctuation}

In this subsection
we study the thermal fluctuation of the bond distortion $\{\Delta_i\}$.
First
we investigate the distribution function of the bond distortion.
In case of the bond alternation
the distribution has two peaks at the corresponding position.
In Figs.~\ref{fig:his}
we show the normalized distribution of the bond distortion
for several values of $k$ at several temperatures.
As the temperature decreases
the wide distribution becomes narrow
and the distribution changes over
from a one-peak distribution to a two-peak distribution.
The position of the peak corresponding to the bond alternation
is consistent with that in the ground state estimated by ED in the previous section.
It should be noted that
the distribution becomes asymmetrical in the intermediate stage to the two-peak distribution.
The amplitude at the peak of positive-$\Delta$ is larger than
that at the peak of negative-$\Delta$.
This behavior is interpreted as follows.
It is easy to form the spin singlet state at the strong bond with positive-$\Delta$
so that the energy is lowered at the strong bond.
On the other hand,
if the spin singlet state is formed at the weak bond with negative-$\Delta$,
the energy is higher than that of the spin singlet pair at the strong bond.
Then the strong bond is more stable than the weak bond.
When the thermal fluctuation of the bond distortion is taken into consideration,
for negative-$\Delta$ the effect of the thermal fluctuation is more remarkable
and the distribution extends to wider range than positive-$\Delta$.
This difference between positive- and negative-$\Delta$ should disappear
in the ground state at $T=0$.
In Fig.~\ref{fig:bmax} we show the temperature dependence of the peak positions
of positive- and negative-$\Delta$ for $k=1.0$.
For negative-$\Delta$ the peak appears at a lower temperature
than that of positive-$\Delta$ as we mentioned above.
Peak positions of positive- and negative-$\Delta$ reach to saturated values
at almost the same temperature.

We find that
the distribution extends to wide range
even at a modestly low temperature $T=0.1$ for $k=2.0$ in Fig.~\ref{fig:his}(b).
At a high temperature $T=0.5$
the distribution extends up to the cutoff ($\Delta = \pm 0.6$).
As shown in Fig.~\ref{fig:his}(a),
the distribution extends to wider range for the case of $k=1.0$.
The cutoff of the distortion restricts
the exchange coupling to the provided range ($J_i=0.4 \sim 1.6$),
which would represent a non-linear effect.
If the cutoff is removed,
it is expected that the distribution extends to ferromagnetic range
and it is not realistic in the spin-Peierls system.
Practically
the non-linearity of the lattice distortion is expected to exist.
Namely,
the linear relation between the lattice distortion and the shift of the exchange coupling
$J_i=J(1+\Delta_i)$ should be changed to include higher orders for large $\Delta$,
or the non-linearity of the elastic energy potential 
$U(x)=\frac{k}{2}x^2+\gamma x^4+\cdots$,
which implies that unreasonable large distortion should be prohibited.
Because the distortion distributes almost uniformly around $\Delta=0$,
it is expected that
the value of the provided cutoff is not essential
to represent the property of the system qualitatively.
Hereafter, we study thermodynamic properties for the case of $k=1.0$,
where the system shows characteristic properties of the spin-Peierls system
at appropriate low temperature.
It is expected that thermodynamic properties such as the magnetic susceptibility
are affected by the thermal fluctuation of the bond distortion at high temperature.

In order to investigate the property of the bond ordering,
we introduce the order parameter
\beq
\Delta_{\rm sg}^2 = \left( {1 \over N}\sum_{i=1}^N (-1)^i \Delta_i \right)^2,
\eeq
which represents how the bond alternates in the whole chain.
From the size dependence of $\Delta_{\rm sg}^2$,
we can estimate the correlation length of the alternate distortion.
\beqa
\Delta_{\rm sg}^2(N)
&=&
\left(\frac{1}{N}\sum_{i}(-1)^i\Delta_i\right)^2 \nonumber \\
&=&
\frac{1}{N^2}\sum_{i,j}(-1)^{i-j}\Delta_i\Delta_j \nonumber \\
&\sim&
\frac{1}{N}\sum_j(-1)^j\Delta_0\Delta_j \nonumber \\
&\sim&
\frac{1}{N}\xi(T),
\eeqa
because $\Delta_0\Delta_j$ decays exponentially
and the summation in the above equation converges to a correlation length $\xi(T)$.
Thus we reach the relation
\beq
\label{eq:cor}
\frac{\Delta_{\rm sg}^2(N)}{\Delta_{\rm sg}^2(N')}=\frac{N'}{N}.
\eeq
In Fig.~\ref{fig:bsg2} we show the temperature dependence of $\Delta_{\rm sg}^2$.
In order to investigate the size dependence,
the values for $N=32$, $64$ and $128$ are compared in the figure.
We find the relation (\ref{eq:cor}) at high temperatures $T>0.1$,
which indicates that the bond correlation is only short range, $\xi \ll N$.
On the other hand,
as the temperature decreases,
the alternate bond distortion develops
near the temperature $T\sim 0.1$
where the distribution becomes to change into the two-peak distribution.
The value of $\Delta_{\rm sg}^2$ of each size
converges to the same value in the ground state,
which indicates that the correlation extends to the whole chain, $\xi \gg N$.
In Figs.~\ref{fig:bcor}
we show explicitly the bond correlation function between the first and the $j$-th bond
\beq
C_{\rm bond}(1,j) = \Delta_1 \Delta_j,
\eeq
at several temperatures.
We find that each bond fluctuates independently
at a high temperature $T=0.5$ as shown in Fig.~\ref{fig:bcor}(a).
As a result, the bond distortion is uniform on average in the whole chain.
As the temperature decreases
the alternate bond correlation develops gradually.
As shown in Fig.~\ref{fig:bcor}(b),
at a modestly low temperature $T=0.1$
short range correlation of the alternate distortion exists.
As shown in Fig.~\ref{fig:bcor}(c),
at a very low temperature $T=0.01$
the alternate bond correlation is spread over the whole chain
and the dimerized ground state is realized.
It should be noted that
the distribution still fluctuates from the complete dimerization.

\subsection{Magnetic susceptibility}

The magnetic susceptibility shows a characteristic feature
due to the spin gap in the spin-Peierls system.
Below the transition temperature the energy gap opens
between the singlet ground state and the triplet excited state
as a result of the bond alternation.
There the susceptibility drops exponentially with decrease of the temperature.
In Fig.~\ref{fig:xmg} we show the temperature dependence of the magnetic susceptibility.
In order to investigate the finite size effect,
the values for $N=32$, $64$ and $128$ are compared in the figure.
The solid line denotes the susceptibility of the uniform chain for $N=128$.
We find that
the susceptibility of each size drops at almost the same temperature,
which indicates that the energy gap causing this drop appears
on account of the bond alternation rather than the finite size effect.
On the other hand,
we find that
the peak position shifts to a lower temperature
and the susceptibility is enhanced at high temperatures
compared to the case of the uniform chain.
As shown in the previous subsection,
the bond distortion fluctuates quite largely at high temperature.
The susceptibility is also affected by the thermal fluctuation of the bond distortion.

It has been pointed out that
the susceptibility of the spin-Peierls system
is different from that of the uniform chain at high temperature,
which is due to the fact that
the uniform phonon displacement causes the effective exchange coupling
larger than the bare exchange coupling.~\cite{ph-1,ph-2}
In these studies
the distortion of each bond behaves independently,
and the total length of the chain is subjected to change.
On the contrary,
in the present case 
we fix the total length of the chain.
Within this restriction
we still found the change of the susceptibility at high temperature.

In order to study the effect of the bond distortion in the present model
on the susceptibility at high temperature,
let us take into account the degree of freedom of the bond.
Namely, we consider the following partition function
\beq
Z=
{\rm Tr} \left( \prod_{i=1}^N \int {\rm d}\Delta_i \right) \,
{\rm e}^{-\beta H} \,
\delta \left( \sum_{i=1}^N \Delta_i \right),
\eeq
where $H$ is given by eq.~(\ref{eq:hamsp}) and
the restriction on the fixed total length of the chain is taking into account
by the delta-function.
We approximate the density matrix by the Suzuki-Trotter decomposition
with the Trotter number $m=1$
\beq
{\rm e}^{-\beta J\sum_i(1+\Delta_i)\mbox{\boldmath$S$}_i\cdot\mbox{\boldmath $S$}_{i+1}}
\cong \prod_i{\rm e}^{-\beta J(1+\Delta_i)\mbox{\boldmath$S$}_i\cdot\mbox{\boldmath $S$}_{i+1}},
\eeq
which would be allowed at high temperature.
We also express the delta-function by the integral formula
\beq
\delta \left( \sum_{i=1}^N \Delta_i \right) =
\int_{-\infty}^{\infty} \frac{{\rm d}q}{2\pi} \, {\rm e}^{{\rm i}q \sum_{i=1}^N \Delta_i}.
\eeq
Here we can integrate out the variables $\{\Delta_i\}$ individually
\beqa
Z_{\rm eff}
&=&
{\rm Tr} \, \int_{-\infty}^{\infty}\frac{{\rm d}q}{2\pi} \, \prod_i
\int_{-\infty}^{\infty}{\rm d}\Delta_i \nonumber \\
&&
\times {\rm e}^{-\beta\left[J(1+\Delta_i)\mbox{\boldmath $S$}_i\cdot\mbox{\boldmath $S$}_{i+1}
                + \frac{k}{2}\Delta_i^2\right]
+ {\rm i}q\Delta_i } \nonumber \\
&=&
{\rm Tr} \, \int_{-\infty}^{\infty}\frac{{\rm d}q}{2\pi} \, \prod_i C \nonumber \\
&&
\times {\rm e}^{\frac{1}{2\beta k}\left({\rm i}q -
                \beta J\mbox{\boldmath $S$}_i\cdot\mbox{\boldmath $S$}_{i+1}\right)^2
- \beta J\mbox{\boldmath $S$}_i\cdot\mbox{\boldmath $S$}_{i+1}},
\eeqa
where the cutoff of $\{\Delta_i\}$ is neglected in order to carry out the Gauss integration.
Applying the Suzuki-Trotter formula with $m=1$ again
and carrying out the integration with respect to $q$,
we have the expression of the partition function
%
%
\beqa
\label{eq:eff-pf}
Z_{\rm eff} &=& C' \, {\rm Tr} \, {\rm e}^{-\beta H_{\rm eff}}, \\
H_{\rm eff} &=&
\left( J+\frac{J^2}{4k} \right) \sum_{i=1}^N
  \mbox{\boldmath $S$}_i\cdot\mbox{\boldmath $S$}_{i+1} \nonumber \\
\label{eq:eff-ham}
&&
+ \frac{J^2}{2kN} \left( \sum_{i=1}^N
  \mbox{\boldmath $S$}_i\cdot\mbox{\boldmath $S$}_{i+1} \right)^2.
\eeqa
In this expression we find that
the effective coupling shifts as $J \rightarrow J+J^2/4k$,
which corresponds that the effective temperature shifts as $T \rightarrow T/(1+J/4k)$.
This decrease of the temperature is inconsistent with the change
of the peak position in Fig.~\ref{fig:xmg}.
The second term
${J^2\over 2kN}\left(\sum_i
\mbox{\boldmath $S$}_i\cdot\mbox{\boldmath $S$}_{i+1} \right)^2$
may cause 
shift of the peak position and enhancement of the susceptibility.
In order to check these points, we investigate the susceptibility
of the effective Hamiltonian (\ref{eq:eff-ham}) by ED.
In Fig.~\ref{fig:eff-xmg}
the temperature dependence of the susceptibility of this model is shown,
which reproduces above points qualitatively.
Thus we conclude that the model of eq.~(\ref{eq:eff-ham})
represents well the present model at high temperature.

As shown in Fig.~\ref{fig:xmg}, we find that
the susceptibilities of the uniform chain and the present model (\ref{eq:hamsp})
agree with each other at fairly high temperature ($T/J>0.8$).
However,
we find a small disagreement at high temperatures in Fig.~\ref{fig:eff-xmg},
although the model of eq.~(\ref{eq:eff-ham}) should be the better
model in the higher temperaure.
This disagreement can be understood as the effect
of the restriction on the range of $\Delta_i$.
In the simulation we restrict the range of the distortion $\Delta_i$,
but here we allow it to change without restriction.
Thus the susceptibility at high temperature depends on the details of the
non-linearity of the lattice.

\section{Inhomogeneity and Lattice Relaxation}

So far we have considered the periodic boundary condition.
We found there that the lattice distortion occurs alternately.
In the uniform Heisenberg model
it has been pointed out that
the inhomogeneity causes
various peculiar magnetic structures.~\cite{imp-EA,imp-L,imp-nishino1}
Here we consider the effect of the lattice distortion in open chains.
It is naturally expected that
the bond configuration changes
in order to minimize the total energy of the spin system and the bond system.
Such lattice relaxation has been studied by a kind of method
where the lattice distortion is determined self-consistently.~\cite{imp-sim1,imp-sim2,imp-sim3}
There a localized structure of a defect of the bond alternation is obtained
and a local staggered magnetization appears around it.
In this section we investigate
how the structures of bond and spin develope at low temperature
by means of the same QMC method as that in the previous section.

In the even chain,
we find that
strong bonds locate at both ends of the lattice,
and the bond alternates regularly as shown in Fig.~\ref{fig:bcor-open}(a).
Moreover, no magnetic structure appears.
In the odd chain,
strong bonds again locate at both ends,
which causes a defect of alternation of the bond configuration.
Namely, in lattices consisting of $4m+1$ sites,
there appears a configuration with two successive weak bonds,
while in lattices with $4m+3$ sites,
a configuration with two successive strong bonds appears,
as shown in Figs.~\ref{fig:odd-chains}.
In these lattices with an odd number of sites,
it is known that
a local magnetic structure appears around the inhomogeneity.~\cite{imp-nishino2}
In Fig.~\ref{fig:bcor-open}(b)
we show the bond configuration for $N=31$ at a very low temperature $T=0.01$.
There the profile is close to $\cos x$,
but not the soliton-like shape.
This sinusoidal shape is explained by
the diffusion of the localized soliton-like shape.~\cite{Villain,M-M-R}
Thus we conclude that
the defect is quite diffusive and
we can not determine the position very locally.
This effect causes the average profile of the magnetization to be also sinusoidal
as shown in Fig.~\ref{fig:susl},
where the local susceptibility $\chi_i$ is plotted.
\beq
\left. \chi_{i} \equiv \frac{\partial}{\partial h} \langle S_i^z \rangle 
\right |_{h=0} = \beta \sum_{j} \langle S_j^z S_i^z \rangle \, .
\eeq
This quantity is proportional to the local magnetization profile.
In Fig.~\ref{fig:suslf}
we also show the local field susceptibility $\chi_i^{\rm local}$ defined as
\beq
\left. \chi_{i}^{\rm local} \equiv \frac{\partial}{\partial h_i} \langle S_i^z\rangle
\right |_{h_i=0} = \int_0^{\beta} {\rm d}\tau \langle S_i^z(\tau) S_i^z(0) \rangle \, ,
\eeq
which indicates the degree of quantum fluctuation at each site.~\cite{R-Y,I-M-R}
We do not see any localized structure, either.

Here we find that the soliton-like local defect is quite diffusive
and the staggered magnetic structure spreads in the open chain,
which may cause enhancement of staggered magnetic order
even at very low concentration of non-magnetic doping in the spin-Peierls material,
namely, very low concentration of the defect.

\section{Summary}

In this paper we investigated thermodynamic properties of
a one-dimensional $S=1/2$ antiferromagnetic Heisenberg model coupled to a lattice distortion
by a Monte Carlo method.
In particular we investigated how the dimerization develops as a function of the temperature.
The thermal fluctuation of the bond distortion was taken into consideration.

By means of calculating the bond correlation function,
we found clear evidence that
the alternate bond correlation develops as the temperature decreases
and the dimerized ground state is realized at low temperature.
On the other hand,
the distortion of each bond fluctuates independently at high temperature,
where the alternate correlation is only short range
and the distortion is uniform on average in the whole chain.
Besides,
we investigated the distribution function of the bond distortion.
As the temperature decreases
the distribution changes over from a one-peak distribution to a two-peak distribution.
The position of the peak corresponding to the bond alternation
is consistent with that of the ground state.
The distribution is asymmetric
at intermediate temperature along the way to the two-peak distribution.
This asymmetrical behavior is interpreted as a quantum effect
that the spin singlet state is easy to form at the strong bond to lower the energy.
We found that the thermal fluctuation of the exchange coupling is
quite large at high temperature.
In the present case
we provided the cutoff of the bond distortion so that
the exchange coupling should be antiferromagnetic.
The distribution would extend to ferromagnetic range if the cutoff is removed.
We considered that
the cutoff represents a kind of the non-linearity,
namely,
the non-linearity in the relation
between the bond distortion and the shift of the exchange coupling,
or the non-linearity of the elastic energy.
Although appearance of ferromagnetic bonds is not realistic in the spin-Peierls system,
properties of such a system are also of interest in a general point of view.

We also investigated the temperature dependence of the magnetic susceptibility.
We found that the susceptibility drops with decrease of the temperature at low temperature,
which is a characteristic feature in the spin-Peierls system.
We also found that the susceptibility is enhanced at high temperature
compared to that of the uniform chain.
In particular
we found the change of the susceptibility 
even in the case where we fix the total length of the lattice.

We considered the inhomogeneity of the lattice
taking the lattice relaxation into consideration.
We investigated the structures of bond and spin in open chains.
We found that
strong bonds locate at the edges both in the even and odd chains,
and a defect of the bond alternation appears in the odd chain.
The soliton-like defect moves quite diffusively
and the bond structure forms a sinusoidal configuration
rather than the localized soliton-like one,
on the contrary
a soliton-like structure is obtained
within a self-consistent treatment of the lattice relaxation.~\cite{imp-sim1,imp-sim2,imp-sim3}
An staggered magnetic structure also spreads over the chain,
which would give an explanation why
very low concentration of doping of non-magnetic impurities causes
enhancement of the staggered magnetic order
in the spin-Peierls material.~\cite{exp-imp2}

\section*{Acknowledgements}

The authors would like to thank
Professor N. Kawashima for his kind guidance to make a computer code of LCQMC,
and acknowledge M. Nishino and P. Roos
for their assistance in further improvement of the computer code.
The authors also thank
Professor H. Nishimori
for the use of the package TITPACK for exact diagonalization.
The present study is partially supported by
Grant-in-Aid from the Ministry of Education, Science, Sports and Culture of Japan.
They also appreciate for the facility of Supercomputer Center,
Institute for Solid State Physics, University of Tokyo.


%
%
\begin{figure}
\caption{
Possible site movements corresponding to the lattice displacement in the Monte Carlo simulation.
The lower figure shows that the site moves to the right for an example.
The bond distortions on both sides of the site are changed according to the site movement.
}
\label{fig:site}
\end{figure}
%
%
\begin{figure}
\caption{
The $\delta$ dependence of the change of the ground state energy $\Delta E_g$,
estimated by ED for $N=20$.
}
\label{fig:dele}
\end{figure}
%
%
\begin{figure}
\caption{
The $k$ dependence of the dimerization amplitude in the ground state $\delta_0$,
estimated by ED for $N=20$.
The inset represents the relation $\delta_0 \propto k^{-3/2}$,
where the solid line is the guide to the eye.
}
\label{fig:del0}
\end{figure}
%
%
\begin{figure}
\caption{
The distribution function of the bond distortion obtained by QMC.
(a)~$k=1.0$ and (b)~$k=2.0$ for $N=128$.
}
\label{fig:his}
\end{figure}
%
%
\begin{figure}
\caption{
The temperature dependence of the peak position of the distribution of the bond distortion
obtained by QMC.
$k=1.0$ and $N=128$.
The symbols open circle and solid circle denote
the peak position in positive and negative range of $\Delta$, respectively.
}
\label{fig:bmax}
\end{figure}
%
%
\begin{figure}
\caption{
The temperature dependence of $\Delta_{\rm sg}^2$ obtained by QMC.
$k=1.0$.
}
\label{fig:bsg2}
\end{figure}
%
%
\begin{figure}
\caption{
The bond correlation function $C_{\rm bond}(1,j)=\Delta_1\Delta_j$
obtained by QMC.
(a) $T=0.5$, (b) $T=0.1$ and (c) $T=0.01$
for $k=1.0$ and $N=128$.
}
\label{fig:bcor}
\end{figure}
%
%
\begin{figure}
\caption{
The temperature dependence of the magnetic susceptibility
obtained by QMC.
$k=1.0$.
The solid line denotes the values of the uniform Heisenberg chain for $N=128$.
}
\label{fig:xmg}
\end{figure}
%
%
\begin{figure}
\caption{
The temperature dependence of the magnetic susceptibility
of the effective Hamiltonian (\ref{eq:eff-ham})
obtained by ED.
$k=1.0$ and $N=12$.
The solid line denotes the values of the uniform chain.
}
\label{fig:eff-xmg}
\end{figure}
%
%
\begin{figure}
\caption{
The bond configuration in open chains obtained by QMC.
(a) $N=32$ and (b) $N=31$ for $k=1.0$ at $T=0.01$.
}
\label{fig:bcor-open}
\end{figure}
%
%
\begin{figure}
\caption{
The conceptual configuration of the bond alternation with a defect
in odd chains with (a) $4m+1$ sites and (b) $4m+3$ sites.
The thick line denotes the strong bond.
}
\label{fig:odd-chains}
\end{figure}
%
%
\begin{figure}
\caption{
The local susceptibility $\chi_i$ obtained by QMC.
For the open chain with $N=31$ for $k=1.0$ at $T=0.01$.
}
\label{fig:susl}
\end{figure}
%
%
\begin{figure}
\caption{
The local field susceptibility $\chi_i^{\rm local}$ obtained by QMC.
For the open chain with $N=31$ for $k=1.0$ at $T=0.01$.
}
\label{fig:suslf}
\end{figure}

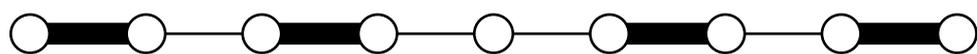
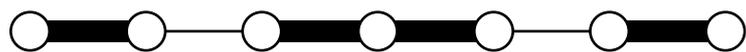
\end{document}